\documentclass[journal=ancac3,manuscript=article]{achemso}

\usepackage{graphicx,textcomp,natmove}

\author{Mukesh Tripathi}
\affiliation{University of Vienna, Faculty of Physics, 1090 Vienna, Austria}
\email{mukesh.tripathi@univie.ac.at}

\author{Alexander Markevich}
\affiliation{University of Nottingham, School of Chemistry, NG7 2RD Nottingham, UK}

\author{Roman B{\"o}ttger}
\author{Stefan Facsko}
\affiliation{Helmholtz-Zentrum Dresden-Rossendorf, Institute of Ion Beam Physics and Materials Research, 01314 Dresden, Germany}

\author{Elena Besley}
\affiliation{University of Nottingham, School of Chemistry, NG7 2RD Nottingham, UK}

\author{Jani Kotakoski}
\author{Toma Susi}
\affiliation{University of Vienna, Faculty of Physics, 1090 Vienna, Austria}
\email{toma.susi@univie.ac.at}

\title{Implanting germanium into graphene}

\begin{document}

\begin{abstract}
Incorporating heteroatoms into the graphene lattice may be used to tailor its electronic, mechanical and chemical properties. Direct substitutions have thus far been limited to incidental Si impurities and P, N and B dopants introduced using low-energy ion implantation. We present here the heaviest impurity to date, namely $^{74}$Ge$^+$ ions implanted into monolayer graphene. Although sample contamination remains an issue, atomic resolution scanning transmission electron microscopy imaging and quantitative image simulations show that Ge can either directly substitute single atoms, bonding to three carbon neighbors in a buckled out-of-plane configuration, or occupy an in-plane position in a divacancy. First principles molecular dynamics provides further atomistic insight into the implantation process, revealing a strong chemical effect that enables implantation below the graphene displacement threshold energy. Our results show that heavy atoms can be implanted into the graphene lattice, pointing a way towards advanced applications such as single-atom catalysis with graphene as the template.
\end{abstract}

\flushbottom
\maketitle
\thispagestyle{empty}


Graphene~\cite{Novoselov20004Science} is an atomically thin single layer of hexagonally bound carbon with remarkable thermal~\cite{Balandin2008Nanolett.}, mechanical~\cite{Lee2008Science}, and electrical properties~\cite{BOLOTIN2008Solidstatecomm.}. This makes it an ideal material for applications ranging from composites to optoelectronics. However, for many purposes, the atomic structure of graphene is chemically too inert. One way to modify this is to incorporate heteroatoms as impurities into the structure, commonly introduced via a suitable chemical precursor during synthesis~\cite{Guo2010Nanolett.}. However, due to a lack of control over their bonding and contamination by different chemical remnants, the quality of the produced structures is often limited~\cite{Banhart2011ACSNano,Ito14AN}. An alternative is to use ion implantation.

Ion implantation is a mature technique to manipulate materials that offers great flexibility: a plethora of ion species, a wide range of implantation energies, and control over the dopant concentration through the ion fluence. It is widely used by the modern semiconductor industry to introduce controlled concentrations of \textit{p} and \textit{n}-type dopants at precise depths below crystal surfaces. Recently, single ions in semiconductors such as silicon, germanium and arsenic have attracted interest for the development of quantum computers~\cite{Persaud2004Quant.infoprocess}. However, ion implantation is quite challenging in the case of two-dimensional (2D) materials such as graphene since only a narrow energy window will allow implantation---high enough to remove one or more target atoms yet low enough to stop the ion within the atomically thin structure~\cite{AAhlgren11PRB}. Moreover, since adatoms on 2D crystals tend to be mobile~\cite{Ossi2015Nanolett} and there are typically no available interstitial sites, the implanted ion needs to form covalent bonds with the under-coordinated atoms of the vacancy to reach a stable configuration. After early work on carbon nanotubes~\cite{Kotakoski05PRB,Xu07SS}, the direct substitutional doping of graphene~\cite{Bangert2013Nanolett,Willke2015Nanolett} and transition metal dichalcogenides~\cite{BANGERT2017Ultramic} has been achieved using low-energy ion implantation, but large-scale controlled dopant incorporation into 2D materials remains an important challenge.

Germanium (Ge) is a structural and electronic analogue to silicon (Si), but significantly heavier (atomic number 32 as compared to 14) and larger (covalent atomic radius of 122 pm as compared to 111 pm for Si and 77 pm for C). This raises the question whether it could also be incorporated into graphene similar to the lighter boron and nitrogen~\cite{Bangert2013Nanolett,Kepaptsoglou2015ACSNano}, the often observed silicon~\cite{Zhou2012PhysRevlett.,Quentin2013Nanolett,Susi17UM}, and the recently implanted phosphorus~\cite{Susi172DM}, all of which are able to directly substitute for single C atoms. Due to its greater size, a significant increase of bond length resulting in threefold Ge buckling out of the plane is predicted~\cite{OuldNe2017optiandquantelec}. Its isovalence with Si anticipates~\cite{Quentin2013Nanolett} another possible bonding: a (nearly~\cite{Nieman17JCP}) planar fourfold  substitution in a double vacancy. Based on density functional theory (DFT) calculations, doping graphene with germanium can open and control the band gap depending on concentration~\cite{Denis2014ChemPhysChem,OuldNe2017optiandquantelec}. However, no experimental studies have yet been reported.

\section*{Results and discussion}
We have successfully incorporated Ge into the graphene lattice using low-energy ion implantation. Although even our best implanted samples remain heavily contaminated~\cite{Susi172DM}, atomic resolution scanning transmission electron microscopy (STEM) shows that Ge can directly substitute for single atoms, bonding to three carbon neighbors in a buckled out-of-plane configuration as predicted by DFT. The chemical nature of the dopants is confirmed through quantitative image simulations and energy dispersive x-ray spectroscopy (EDX), whereas electron energy loss spectroscopy (EELS) is hardly sensitive to Ge due to spectral overlap with the much more intense graphene $\pi+\sigma$ plasmon response~\cite{Eberlein08PRB}. We describe the atomic-level details of the implantation process through \textit{ab initio} molecular dynamics (MD) simulations, which reveal a strong chemical effect that reduces the required implantation energy for Ge as compared to purely kinematic arguments. This is experimentally verified by successful implantation below the carbon displacement threshold energy in graphene. Our results show for the first time that elements beyond the third period can be directly implanted into graphene, expanding the versatility of ion irradiation as a tool for engineering graphene.

\subsection*{Experimental}
\begin{figure}[t!]
\centering
\includegraphics[width=\linewidth]{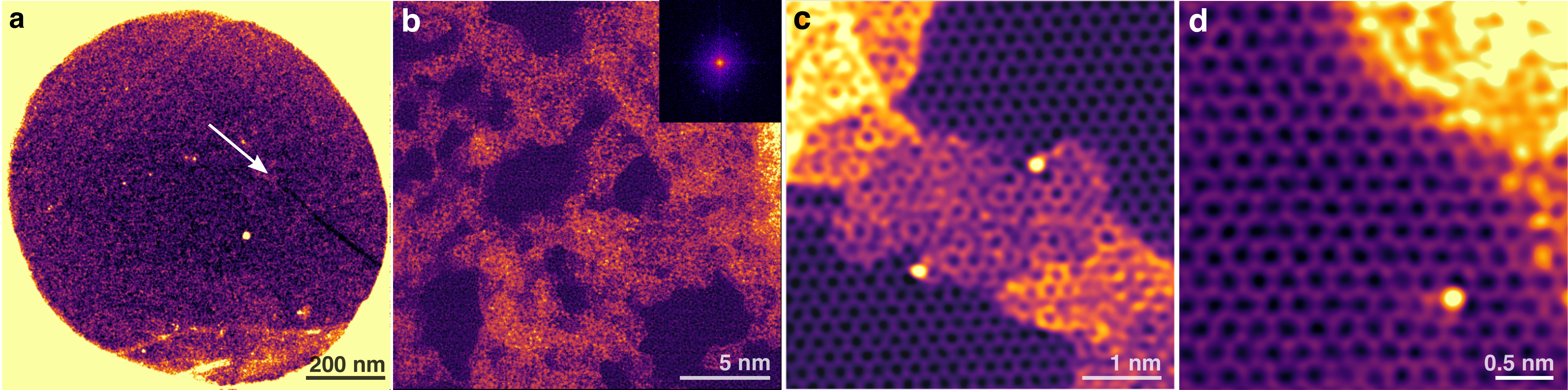}
\caption{{\bf Overviews of a graphene sample implanted with 20 eV $^{74}$Ge$^+$ ions} (MAADF/STEM, $1024 \times 1024$ px). (a) Raw image of heavily contaminated graphene suspended over a Quantifoil hole after ion implantation, with the arrow indicating a narrow cleaner region. (b) Clean monolayer graphene areas some tens of nm$^{2}$ in size were found in this region. The fast Fourier transform (FFT) of the inset shows the hexagonal structure of the graphene lattice. (c) Two Ge atoms bonded at the edge of overlaying graphitized contamination. (d) A single Ge atom incorporated into the graphene lattice. Panels b--d have been processed with double Gaussian filtering to reduce the probe tail effect~\cite{Krivanek2010nature}.}
\label{fig:overviews}
\end{figure}

Contamination introduced during implantation degrades sample quality and makes characterization difficult. Like in previous P-implanted samples~\cite{Susi172DM}, most of the graphene surface is covered by a thick layer of contamination (a medium angle annular dark field, MAADF, image is presented in Fig.~\ref{fig:overviews}a), despite laser treatment both before and during the implantation (Methods). The degree of contamination was however not entirely consistent: in some samples we were not able to find any clean lattice despite extensive searching, whereas in others this was  rare, but included regions with implanted impurity atoms. Due to this variation, which was greater than in earlier P-implanted graphene~\cite{Susi172DM}, we cannot exclude the possibility that our laser alignment varied between the different experiments.

Most of our data is from the cleanest sample that was irradiated at 20~eV. We did also find one incorporated impurity atom in a sample irradiated at 15~eV, but found no clean lattice in the sample irradiated at 25~eV. This lack of statistics from different ion energies prevents us from deducing the ideal implantation energy from the experimental data---in total, we found only seven Ge impurities directly in the lattice (roughly 1500 nm$^2$ of clean lattice was analyzed; with our ion fluence of 1 ion/nm$^2$, this yields an implantation efficiency of just 0.5\%). An example of a relatively clean sample area is presented in Fig.~\ref{fig:overviews}b, illustrating how most clean areas do not contain any Ge atoms (or other defects). Examples of heavy impurities found in the contamination as well as incorporated into the lattice are presented in Figs.~\ref{fig:overviews}c,d.

\begin{figure}[t!]
\centering
\includegraphics[width=0.9\linewidth]{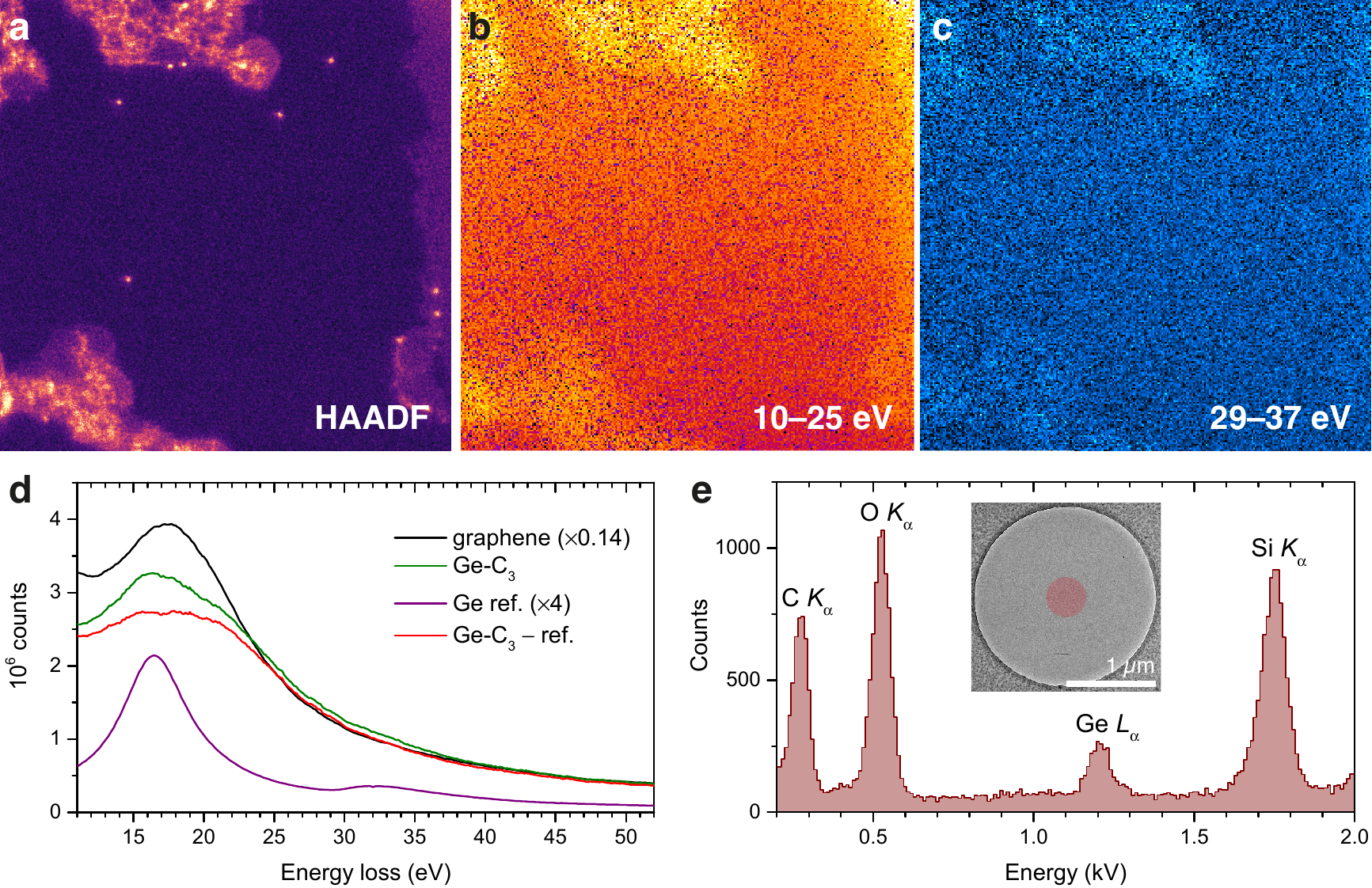}
\caption{{\bf Spectroscopic characterization of implanted germanium.} (a) HAADF/STEM image of a region with multiple Ge atoms found in the lattice and in the contamination layer. (b) Background-subtracted EELS spectrum map integrated over the $\pi+\sigma$ plasmon response. (c) As in b, but integrated over the Ge \textit{M}$_{4,5}$-edge. (d) Spot EELS spectra measured over pristine graphene and over a single Ge substitution (Ge-C$_3$, shorter exposure), along with a thin-film Ge reference (Ge ref.) and their difference (Ge-C$_3 - $ref.). (e) EDX spectrum with the approximate beam size indicated by the colored red area in the inset bright-field image of graphene suspended over a hole in the carbon support foil.}
\label{fig:spectroscopy}
\end{figure}

EELS is usually the best tool for establishing chemical identities and bonding at the atomic level~\cite{Bangert2013Nanolett,Kepaptsoglou2015ACSNano,Susi172DM}. However, Ge impurities in graphene present a particular challenge: the weak Ge \textit{L}-edge is located at $\sim$1150 eV, beyond the range of our spectrometer, whereas the \textit{M}$_{4,5}$-edge, located at $\sim$32 eV, falls under the much more intense graphene $\pi+\sigma$ plasmon (which is further influenced by nearby contamination~\cite{Zhou12NN,Hage18AN}). As a consequence, in the EEL spectrum maps of Figs.~\ref{fig:spectroscopy}b,c, it is not possible to distinguish the signal of the Ge atoms from the background of the plasmon tail, even though the impurities are clearly visible in the simultaneously acquired high angle annular dark field (HAADF) image (Fig.~\ref{fig:spectroscopy}a). Nonetheless, as shown in Fig.~\ref{fig:spectroscopy}d, a high signal-to-noise spot spectrum recorded over a single Ge substitution shows an apparent splitting of the plasmon peak as well as additional intensity compared to pristine graphene at higher energy losses. Although subtracting a scaled reference spectrum of thin-film bulk Ge (Gatan EELS Atlas~\cite{GatanAtlas}) with an intense plasmon component centered at $\sim$17 eV in addition to the \textit{M}$_{4,5}$-edge starting at 29 eV (Ref.~\citenum{Nguyen12PRB}) can partially account for both features, further studies with monochromated EELS~\cite{Krivanek14N} seem warranted. However, as shown in Fig.~\ref{fig:spectroscopy}e, it is much easier to detect the spectroscopic signature of Ge using EDX, which should in a suitable instrument be feasible down to the single-atom level~\cite{Lovejoy12APL}.

\begin{figure}[t!]
\centering
\includegraphics[width=1\linewidth]{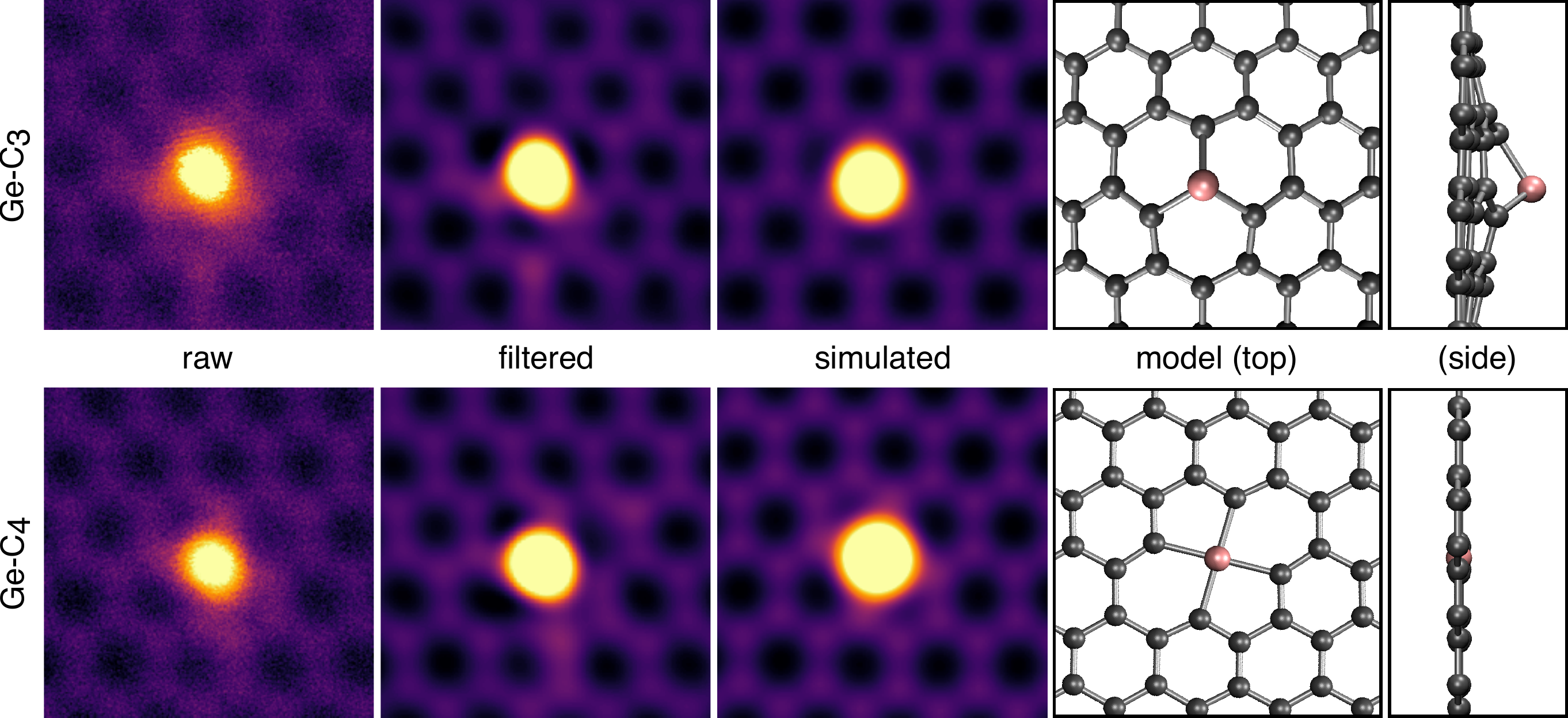}
\caption{{\bf Germanium substitutions in graphene} (top:  threefold single atom substitution, Ge-C$_3$; bottom: fourfold substitution in a double vacancy, Ge-C$_4$). The raw MAADF/STEM images have been averaged from 7 and 40 frames, and further double Gaussian filtered. Quantitative STEM simulation using our experimental parameters reproduces in both cases the high intensity of the Ge impurity. DFT simulations show that while the Ge-C$_3$ impurity buckles out of the graphene plane, the Ge-C$_4$ site is flat. The fields of view are $\sim$1$\times$1 nm$^2$.}
\label{fig:QSTEM} 	
\end{figure}

Due to the atomic number-dependent scattering contrast in annular dark field images~\cite{Krivanek2010nature}, we can turn to quantitative image simulations to verify the identity of the observed impurities~\cite{Susi17SR}. The atomic resolution STEM images shown in Fig.~\ref{fig:QSTEM} give two examples of an individual heavy impurity within the graphene lattice. To reduce the influence of the electron beam probe tails, these MAADF images have been processed using a double Gaussian filter~\cite{Krivanek2010nature}. Scattering at the impurity is extremely intense even to the MAADF detector, making it difficult to distinguish the local structure. However, simulated images of threefold (Ge-C$_3$) and fourfold (Ge-C$_4$) germanium substitutions (Fig.~\ref{fig:QSTEM}) agree well with the filtered images. From HAADF images (that avoid nonlinear scattering effects present in MAADF images), we measure the ratio of the intensity of the impurity atom to that of carbon atoms distant from it as 21.5$\pm$0.9 for Ge-C$_3$ and 21.7$\pm$1.0 for Ge-C$_4$, with simulated ratios respectively of 21.8 and 22.5. This corresponds to a Ge/C contrast proportional to Z${^{1.86}}$, and the agreement confirms the chemical identity of the impurities. Our DFT simulation reproduces the significant structural rearrangement expected around the Ge-C$_3$ site~\cite{OuldNe2017optiandquantelec}, with the Ge atom buckling 1.90 \AA\ out of the graphene plane as shown in the top and side views of the relaxed model structure (Fig.~\ref{fig:QSTEM}).

\begin{figure*}[t!]
\includegraphics[width=0.75\textwidth]{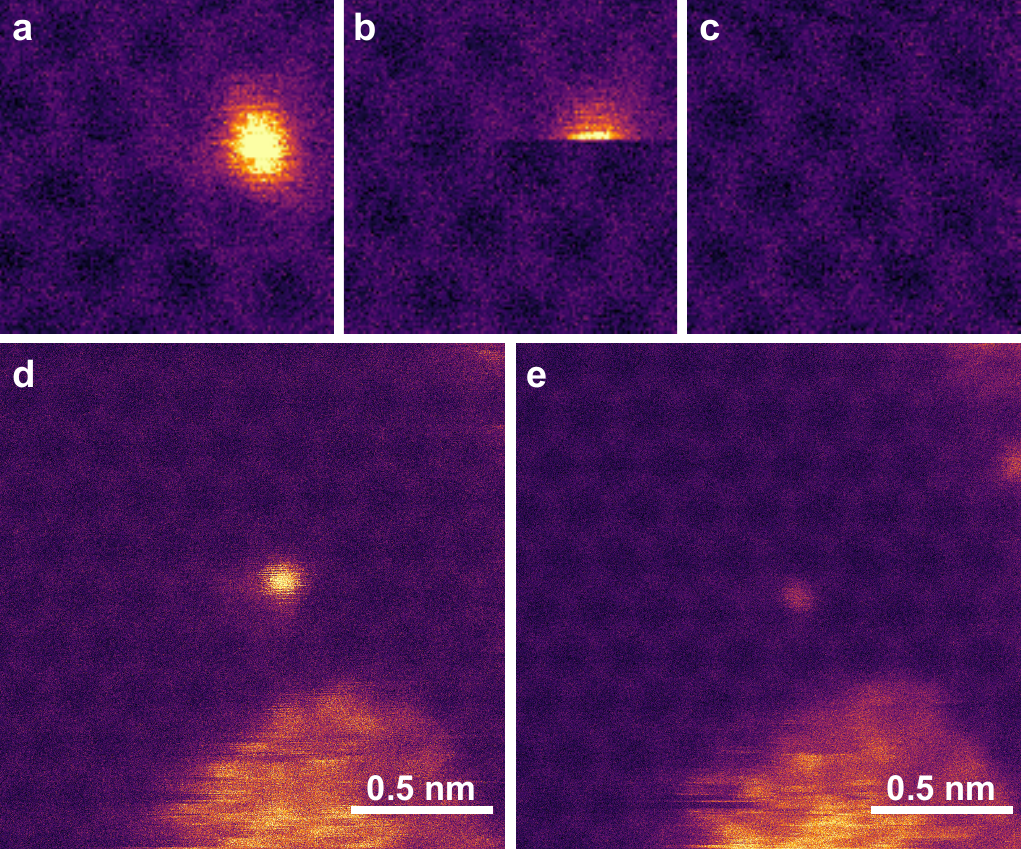}%
\caption{\textbf{Ge replacement under electron irradiation} (MAADF/STEM). a-c) Three consecutive raw images of a single Ge impurity being replaced by C during the second frame ($\sim$4 s per frame). d-e) Replacement of another Ge impurity by Si.\label{fig:replacement}}
\end{figure*}

Finally, inspired by recent electron-beam manipulation of impurity atoms in graphene~\cite{Susi17UM,Susi172DM}, we attempted to move the Ge atoms by iteratively placing the electron beam on one C neighbor for 10 s between acquiring images. A total of 24 such irradiations were attempted on several different impurities, without success. DFT/MD simulations (described below) confirmed the reason: the beam-induced out-of-plane dynamics of the C atom, all the way up to its knock-on threshold of 15.75~eV, are unable to trigger the bond inversion mechanism~\cite{Susi14PRL} for Ge due to its greater mass and covalent radius. For Ge itself, the threshold energy is 11.75~eV, almost an order of magnitude more than what a 60-keV electron can transfer to such a heavy nucleus. However, we observed a curious effect: several Ge impurities got replaced by C (Fig.~\ref{fig:replacement}a-c) or, in one case, Si (Fig.~\ref{fig:replacement}d-e). Presumably the source of these atoms is the ubiquitous contamination, but the electron beam must somehow be involved in the mechanism (since otherwise we would not be able to find any Ge in the lattice).

\subsection*{Modeling}
DFT-based molecular dynamics simulations were performed to analyse the energetics and atomic scale details of the implantation process. For successful incorporation of a heteroatom into graphene lattice by ion irradiation two conditions need to be satisfied: i) the energy transferred from the incident ion to carbon atom during collision is large enough to knock out carbon(s), and ii) the remaining kinetic energy of the ion after collision is low enough so that it stays in the structure. The energy transferred from the incident ion to a lattice atom is maximum for the case of a head-on collision, and therefore such impacts are most efficient in satisfying the above conditions. We observed a number of outcomes depending on the kinetic energy of the incident Ge ion. Ions with energies below 20~eV typically cannot induce breaking of the C-C bonds and thus remain on the graphene surface as adatoms. For ion energies between 20--22~eV, Ge substitutes carbon atom in the lattice; however, the displaced carbon is not sputtered but remains bound to the system. All our simulations show that such configurations relax spontaneously to a structure consisting of non-defective graphene with a Ge adatom (although this may not be the case in experiments at room temperature). For Ge energies of 23--25~eV, a Ge-C dimer is formed that flies away from the graphene layer, in which a monovacancy defect is formed. Finally, at 26~eV Ge is directly implanted into graphene structure while the C atom is sputtered. According to our calculations, direct implantation occurs for Ge kinetic energies ranging from 26 to 42~eV, while ions with higher energy penetrate through the graphene layer. 

We find that in a head-on collision, an incident Ge ion transfers about 49\% of its initial kinetic energy to an isolated carbon atom. The maximum kinetic energy of the impacted C atom in graphene is slightly less, 10.7 eV (41\%) for an ion energy of 26 eV (Fig. 5). This is significantly lower than the threshold energy for displacement of carbon in graphene, calculated with our method as 22.3 eV. Evidently, interactions between the incident Ge ion and C atoms in graphene significantly lower the displacement threshold energy of carbon.

\begin{figure}[th!]
\centering
\includegraphics[width=0.65\linewidth]{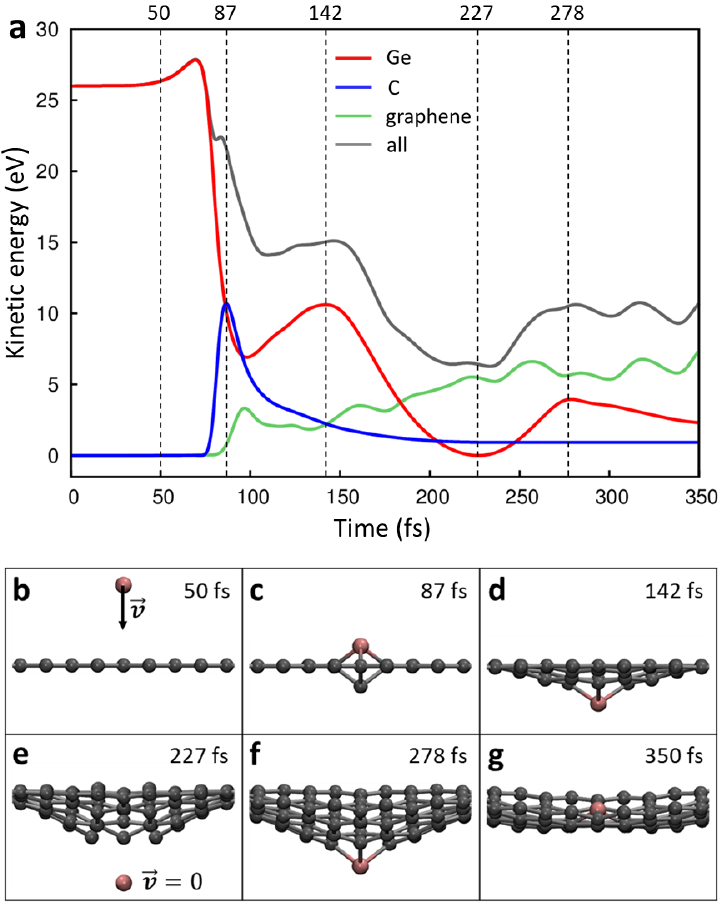}
\caption{\textbf{Energetics and atomic scale details of the implantation process.} (a) Kinetic energies of the incident Ge ion (shown in red colour), sputtered C atom (blue), all remaining C atoms in the graphene layer (green) and total kinetic energy of the system (grey) as a function of time, obtained from the DFT/MD calculations for an initial ion energy of 26 eV. b-g) Atomic configurations that correspond to time steps marked in (a) by vertical dashed lines. The Ge atom is shown with pink colour, while carbon atoms are shown in grey.}
\label{fig:Theory} 	
\end{figure}

To obtain a more detailed picture of the process, we analyse the kinetic energies of the incident Ge ion ($T_\textrm{Ge}$), sputtered C atom ($T_\textrm{C}$), remaining C atoms in the graphene layer ($T_\textrm{gr}$) and the total kinetic energy of the system ($T_\textrm{tot}$) at each time step of the simulation for an initial ion energy of 26~eV (Fig.~\ref{fig:Theory}). At the beginning of the simulation, the total kinetic energy of the system equals that of the Ge ion. It slightly increases when the ion approaches graphene, with a maximum value of 27.8~eV, and starts to decrease quickly when the separation between the ion and graphene plane becomes less than 2.2~\AA. The shortest Ge-C distance was found to be 1.53~\AA. The energy transfer between the incident ion and the carbon atom occurs over tens of fs. When its kinetic energy reaches its maximum value of 10.7~eV, the C is already displaced from graphene plane (Fig.~\ref{fig:Theory}c) and, therefore, the maximum of $T_\textrm{C}$ can be lower than the actual amount of the transferred energy. Subsequently, the kinetic energy of the sputtered C atom drops quickly by about 6~eV and then gradually decreases to a constant value of 0.92~eV.

Interestingly, in simulations with higher energies, the long tail after the initial drop of $T_\textrm{C}$ starts to disappear. For example, for an initial Ge energy of 40 eV, $T_\textrm{C}$ becomes constant within 40 fs and decreases by only 5.7 eV from the maximum. This suggests that the C needs a kinetic energy of only about 5.7 eV to overcome the interaction with the graphene lattice after a head-on collision with Ge. The long tail in $T_\textrm{C}$ at low ion energies occurs mainly because of the interaction of C with the Ge ion, which can be considerable when both atoms are in close proximity and on the same side of graphene plane. At high values for the initial energies of the incident ion, the knocked-out C atom quickly moves away from graphene before the ion passes through the layer and therefore the tail in $T_\textrm{C}$ disappears.

Coming back to the example shown in Fig.~\ref{fig:Theory}, the Ge ion continues to move in the same direction after the collision and the amount of kinetic energy transferred to C atoms in the lattice increases. A minimum in $T_\textrm{Ge}$ at around 100 fs corresponds to the point when the ion passes through the graphene layer. The subsequent increase in $T_\textrm{Ge}$ occurs partially because the ion is moving towards a more energetically favourable configuration (Fig.~\ref{fig:Theory}d) and partially due to the interaction with the sputtered C atom. The energy increase is significantly smaller in simulations for higher ion energies, confirming that at low ion energies the Ge-C interaction plays an important role. The kinetic energy of the ion then goes to zero while $T_\textrm{gr}$ gradually increases. Figure~\ref{fig:Theory}e shows the configuration at which $T_\textrm{Ge}$ equals zero. Although the ion has travelled a significant distance (about 4.3~\AA) past the plane of graphene, the interaction with the highly buckled lattice is strong enough to stop and reverse the motion of the ion. The distance between the Ge and the undercoordinated C atoms at this time step is about 2.9~\AA. Moving backwards, the ion gains enough kinetic energy to pass again through the graphene layer (Fig.~\ref{fig:Theory}f-g), before it finally stops, bound to the lattice. Further atomic motions in the system are relatively slow and do not involve significant structural rearrangements. 

Finally, we should address the success of experimental implantation at energies below those predicted by our modeling. Considering the finite precision of the experimental ion energy, possible variation in the local bias potential, and the sources of inaccuracy in the simulations (including the approximation of exchange and correlation, the neglect of spin, and any basis set superposition error), as well as the simulated penetration of Ge into the lattice already between 20--22 eV, the modeling is actually surprisingly accurate. A statistical comparison of different experimental ion energies, as well as simulations covering the entire impact parameter space, would be required to make more precise direct comparisons.

\section*{Conclusion}
We have implanted thus far the heaviest directly observed graphene impurity, and shown that despite its size, germanium can substitute a single atom by bonding to three carbon neighbors. Our first principles molecular dynamics simulations reveal that due to a strong chemical interaction between the incoming Ge ion and the neighbours of the displaced C atom, implantation can be carried out at an energy below the displacement threshold for graphene. As in earlier works, sample contamination remains an issue, and preventing it during implantation is an important target for improving sample quality in future experiments. Nevertheless, ion implantation continues to progress towards its promise as a scalable and precise technique for controllably doping 2D materials.

\begin{acknowledgement}
M.T. and T.S. acknowledge funding by the Austrian Science Fund (FWF) via project P 28322-N36 and J.K. via project I3181-N36. J.K. was also supported by the Wiener Wissenschafts-Forschungs- und Technologiefonds (WWTF) via project MA14-009. A.M. and E.B. acknowledge funding from the European Research Council under the European Union's Seventh Framework Programme (FP7) / ERC grant agreement \textnumero~307755-FIN. We thank the technical staff of the ion beam center facility of HZDR Dresden for providing us with support and facilities for the ion implantation experiments. DFT-based molecular dynamics simulations were performed using the high performance computing facility at the University of Nottingham, and further simulations at the HPC Midlands Plus Centre and the Vienna Scientific Cluster (VSC).
\end{acknowledgement}

\section*{Methods}
Low-energy $^{74}$Ge$^+$ ions were implanted at the HZDR ion beam center into commercially available monolayer graphene supported on Au TEM grids (Quantifoil\textcircled{R} R 2/4, Graphenea). We used a mass-selected, twofold electrostatic raster-scanned ion implantation system (Danfysik A/S, Denmark, Model 1050), providing ion energies down to 100 eV. Neutralized ions were eliminated by deflecting the $^{74}$Ge$^+$ ion beam and then decelerating it towards the target. To reduce energies down to 15 eV, a bias voltage was set to the sample holder by an adjustable anode potential. The samples were irradiated in a $9\times$10$^{-7}$ mbar vacuum at room temperature with a fluence of $1\times$10$^{14}$ cm$^{-2}$ (estimated with multiple Faraday cups). We chose ion energies of 15, 20 and 25~eV to minimize irradiation-induced damage. The amount of energy that can be transferred to a carbon atom due to a collision with such ions is below the experimentally estimated displacement threshold energy (21.14~eV) of graphene~\cite{Susi16NC}. Lower energies than expected are sufficient due to a strong chemical effect between the incoming Ge ion and the structure around the C atom being displaced. In an effort to reduce contamination~\cite{Mukesh2017PSSR}, a 445 nm laser diode was aimed at the sample through a viewport of the vacuum chamber both for 2 min before implantation (nominal laser power 480 mW) and during the implantation (240 mW).

The samples were imaged in near ultra-high vacuum ($< 10^{-9}$~mbar) in an aberration-corrected Nion UltraSTEM100 scanning transmission electron microscope~\cite{Krivanek08UM} operated at an acceleration voltage of 60 kV, well below the knock-on damage threshold of graphene~\cite{Susi16NC}. The beam current was around 50 pA, the beam convergence semi-angle was 30 mrad, and angular range was 60--200 mrad for MAADF and 80--300 mrad for HAADF. Some images were processed using a double Gaussian filtering procedure~\cite{KRIVANEK2010Ultramic} and all colored with the ImageJ lookup table "mpl-magma" to highlight relevant details. Low-loss EELS was recorded in the same instrument using a Gatan PEELS~666 spectrometer retrofitted with an Andor iXon~897 electron-multiplying charge-coupled device (EMCCD) camera~\cite{Susi172DM}. The energy dispersion was 0.1~eV/pixel (with an instrumental broadening of $\sim$0.4~eV) and the EELS collection semiangle was 35~mrad. Additionally, for greater sensitivity for Ge, we collected energy-dispersive x-ray spectra in a Philips CM200 TEM instrument operated at 80 kV.

To model the Ge substitution, we replaced one C atom in a 7$\times$4 orthorhombic supercell of graphene (112 atoms in total), and relaxed its structure via DFT using the GPAW package~\cite{Enkovaara2010} (PBE functional~\cite{Perdew96PRL}, 0.16~\AA\ grid spacing, 5$\times$5$\times$1 Monkhorst-Pack k-points~\cite{Monkhorst76PRB}). The structure was used as input for a quantitative STEM simulation using the QSTEM software package~\cite{Koch2002}, with scattering potentials generated from the independent atom model~\cite{Rez94ACA}. The implantation of Ge was simulated using DFT-based molecular dynamics (MD) as implemented in the SIESTA code~\cite{Soler02JPCM}. We used Troullier-Martins norm-conserving pseudopotentials~\cite{Troullier91PRB}, the PBE functional, and a double-$\zeta$ polarized basis set. The charge density was represented on a real-space grid with an energy cutoff of 300 Ry. For these simulations, graphene was modeled using orthorhombic supercells consisting of 160 carbon atoms. A vacuum layer of 20 \AA\ was included in the direction normal to the graphene plane and the Brillouin zone was sampled using 3$\times$3$\times$1 k-points. To account for dispersion interactions, the Grimme semiempirical potential~\cite{Grimme06JCC} was used. The time step in MD calculations was set to 0.5 fs and the initial kinetic energy of the Ge was varied at 1 eV intervals. Although we call them ions for convenience, note that the projectiles in the simulations are neutral atoms.

\providecommand{\latin}[1]{#1}
\makeatletter
\providecommand{\doi}
  {\begingroup\let\do\@makeother\dospecials
  \catcode`\{=1 \catcode`\}=2\doi@aux}
\providecommand{\doi@aux}[1]{\endgroup\texttt{#1}}
\makeatother
\providecommand*\mcitethebibliography{\thebibliography}
\csname @ifundefined\endcsname{endmcitethebibliography}
  {\let\endmcitethebibliography\endthebibliography}{}

\end{document}